\documentstyle[psfig,graphicx]{mn}                       
\newcommand{\reduceme}{\mbox{R\raisebox{-0.35ex}{E}D%
\hspace{-0.05em}\raisebox{0.85ex}{uc}\hspace{-0.90em}%
\raisebox{-.35ex}{{m}}\hspace{0.05em}E}}

\title[Near-IR line-strengths in elliptical galaxies: evidence for IMF variations?]
{Near-IR line-strengths in elliptical galaxies: evidence for IMF variations?}

\author[A.J. Cenarro et al.]  
    {A.J.~Cenarro,$^1$
    J.~Gorgas,$^1$ A.~Vazdekis,$^2$ N.~Cardiel,$^{1,3}$ and
    R.F.~Peletier$^{4,5}$\\ $^1$Depto. de Astrof\'{\i}sica, Fac. de
    Ciencias F\'{\i}sicas, Universidad Complutense de Madrid, E-28040
    Madrid, Spain\\$^2$Instituto de Astrof\'{\i}sica de Canarias,
    E-38200, La Laguna, Tenerife, Spain\\$^3$Calar Alto Observatory,
    CAHA, Apdo. 511, 04004 Almer\'{\i}a, Spain\\$^4$School of Physics
    and Astronomy, University of Nottingham, University Park,
    Nottingham NG7 2RD, UK\\$^5$CRAL, Observatoire de Lyon, 9,
    Av. Charles Andr\'e, 69230 Saint-Genis-Laval, France\\}
\date{Accepted 2002 December 18. Received 2002 September 20}
\pagerange{\pageref{firstpage}--\pageref{lastpage}}
\pubyear{2000}

\def\LaTeX{L\kern-.36em\raise.3ex\hbox{a}\kern-.15em
    T\kern-.1667em\lower.7ex\hbox{E}\kern-.125emX}

\begin{document}

\label{firstpage}

\maketitle

\begin{abstract}
We present new relations between recently defined line-strength
indices in the near-IR (CaT$^*$, CaT, PaT, MgI, and sTiO) and central
velocity dispersion ($\sigma_0$) for a sample of 35 early-type
galaxies, showing evidence for significant anti-correlations between
Ca\,{\sc ii} triplet indices (CaT$^{*}$ and CaT) and $\log
\sigma_0$. These relations are interpreted in the light of our recent
evolutionary synthesis model predictions, suggesting the existence of
important Ca underabundances with respect to Fe and/or an increase of
the dwarf to giant stars ratio along the mass sequence of elliptical
galaxies.
\end{abstract}

\begin{keywords}
galaxies: elliptical and lenticular -- galaxies: stellar content
\end{keywords}

\section{Introduction}

During the last decade, the measurement and interpretation of blue
optical line-strength indices in the spectra of early-type galaxies in
the field have revealed the existence of an apparent spread of mean
ages (Gonz\'{a}lez 1993; Faber et al. 1995; J{\o}rgensen 1999) and
element abundances ratios (Worthey 1998; Trager et al. 2000a),
suggesting a variety of interpretations of scaling relations like the
colour--magnitude or Mg$_2$--$\sigma$ relations (Bender, Burstein \&
Faber 1993; Kuntschner 2000; Trager et al. 2000b; Vazdekis et
al. 2001). Since the picture from the blue is rather confused, if one
wants to achieve a more complete understanding of the star formation
history of these galaxies, it is necessary to look at other spectral
regions in which the relative contribution of the distinct stellar
types is very different. In this sense, the potential of the near-IR
spectral range and, in particular, of the Ca\,{\sc ii} triplet is
still almost unexploited.

Since Ca is an $\alpha$-element like Mg, it should be enhanced
compared to Fe in giant ellipticals (Es). However, as suggested by
several authors, Ca seems to follow Fe (O'Connell 1976; Vazdekis et
al. 1997; Worthey 1998; Moll\'{a} \& Garc\'{\i}a-Vargas 2000; Vazdekis
et al. 2001; Proctor \& Sansom 2002). Even so, given that variations
of the Fe line-strengths among Es are not negligible (e.g. Gorgas,
Efstathiou \& Arag\'{o}n-Salamanca 1990; Gonz\'{a}lez 1993; Davies,
Sadler \& Peletier 1993; Kuntschner 2000), one should not expect the
small variation of the Ca\,{\sc ii} triplet strength reported by
previous work (Cohen 1979; Bica \& Alloin 1987; Terlevich et al 1990;
Houdashelt 1995). Furthermore, this result is difficult to interpret
in the light of previous stellar population models (Garc\'{\i}a-Vargas
et al. 1998; Schiavon et al. 2000) which predict a high sensitivity of
the Ca\,{\sc ii} triplet to the metallicity of old, metal-rich stellar
populations. Also, the absolute values of Ca\,{\sc ii} in Es differ
from the model predictions (Peletier et al. 1999; Moll\'{a} \&
Garc\'{\i}a-Vargas 2000).

With the aim of clarifying the above inconsistencies, during the last
years we have developed a new stellar library in the near-IR spectral
range (Cenarro et al. 2001a, hereafter CEN01) with a homogeneous set
of revised atmospheric parameters for the library stars (Cenarro et
al. 2001b), deriving empirical fitting functions that describe the
behaviour of new line-strength indices for the Ca\,{\sc ii} triplet
and the H Paschen series (CEN01; Cenarro et al. 2002) and other
spectral features (Cenarro 2002, hereafter CEN02). Finally, in
Vazdekis et al. (2002; hereafter VAZ02) we present a new evolutionary
stellar population synthesis model which predicts both the integrated
indices and the spectral energy distribution for single stellar
populations (SSPs) of several ages, metallicities and initial mass
functions (IMFs).

In this letter we present the first results for a spectroscopic sample
of 35 early-type galaxies. After a brief description of observations
and data reduction (Section~\ref{stellib}), in
Section~\ref{indexsigma} we describe the measurements of the new
indices for the central regions of the galaxies and their relationship
with the velocity dispersion. In Section~\ref{discussion} we discuss
plausible interpretations of the data on the basis of new index--index
diagrams derived from our model predictions.

\section{Observations and data reduction}
\label{stellib}

Our sample consists of 35 early-type galaxies (E -- S0) spanning a
wide range of absolute magnitudes ($-22.5 < M_{\rm B} < -16.5$\,mag,
using H$_0 = 75$\,km\,s$^{-1}$\,Mpc$^{-1}$) and central velocity
dispersions ($40 \la \sigma_0 \la 370$\,km\,s$^{-1}$). Most of them are
field Es, although a few galaxies from Virgo (9) and one cD in the
Coma cluster are also included.

Long-slit spectroscopy was carried out during three nights in 1999
using ISIS at the 4.2\,m William Herschel Telescope (Observatorio del
Roque de los Muchachos, La Palma), providing 2.9\,\AA\ (FWHM) spectral
resolution in the red arm (8355 -- 9164\,\AA). The slit (2\,arcsec
width) was aligned with the major axis except for two S0s (along the
minor axis). Exposure times of 1200 -- 2000\,s per galaxy allowed us
to obtain signal-to-noise ratios per angstrom from 43 to
253\,\AA$^{-1}$ in the central spectra. We followed a typical
spectroscopic reduction procedure with \reduceme\, (Cardiel 1999, see
also CEN01), taking special care on the sky subtraction and the
correction for fringing and telluric absorptions. The availability of
error spectra for each galaxy frame allowed us to estimate reliable
uncertainties in the measurements of the indices. The spectra were
relative-flux calibrated using 4 spectro-photometric standard stars
(Oke 1990) observed several times at different air masses. Also, in
order to correct for small differences between the spectro-photometric
systems of the galaxies and the model predictions, a sample of 49
stars (from B to late M spectral types) in common with 
CEN01 were observed during twilights. They were
also employed as templates for velocity dispersion determinations.

\section{Index--$\log \sigma_0$ relations}
\label{indexsigma}

In this section we present the behaviour of new line-strength indices
in the near-IR spectral region (CaT*, CaT, PaT, sTiO and MgI) as a
function of the central velocity dispersion ($\sigma_0$) of the galaxy
sample.

\begin{figure}
\centerline{\hbox{
\psfig{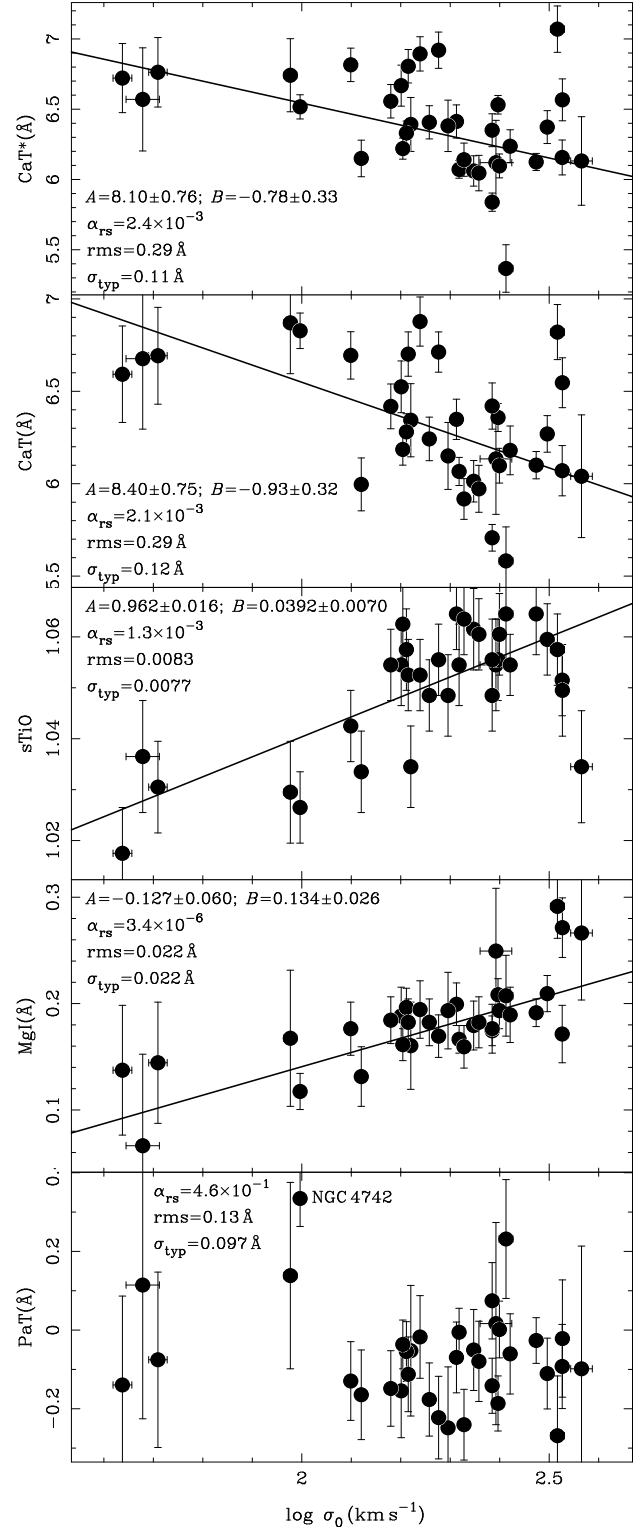}
}} 
\caption{\small 
Behaviour of the new indices with the central velocity dipersion
($\sigma_0$) for the galay sample, at $\sigma = 370$\,km\,s$^{-1}$
spectral resolution and corrected to the system defined by the
models. The lines represent error-weighted, least-squares fits of the
form ${\cal I} = A + B\,\log \sigma_0$ to all data. Labels include the
derived coefficients and their errors, the significance levels of
Spearman rank-order tests ($\alpha_{r_{\rm S}}$), the residual
standard deviation of the fit (rms) and typical index errors for the
galaxy sample ($\sigma_{\rm typ}$). In the case of PaT, rms refers to
the standard deviation w.r.t. the mean index value.}
\label{I-sigma}
\end{figure}

CaT and PaT measure the strength of the Ca\,{\sc ii} triplet
($\lambda\lambda 8498, 8542, 8662$\,\AA) and three lines of the H
Paschen series. CaT$^*$ ($\equiv$ CaT -- 0.93\,PaT) is an index
corrected for the contamination by the Paschen series in stars of the
earliest spectral types (see CEN01 for full details about their
definitions). The sTiO index is a measurement of the slope of the
continuum at the Ca\,{\sc ii} region. It is mainly governed by TiO
absorption bands which are prominent in mid-late M types and observed
in the integrated spectra of early-type galaxies. Following CEN02, it
is computed as the ratio between the CaT* pseudo-continuum values
($C(\lambda)$) at the central wavelength of its reddest and bluest
continuum bands, that is, sTiO $=
C(\lambda8784.0)/C(\lambda8479.0)$. Finally, MgI measures the strength
of the Mg\,{\sc i} line at $\lambda 8807$\,\AA\ (Cenarro et al. 2001c;
CEN02).

Central velocity dispersions for the galaxies were determined using
the MOVEL and OPTEMA algorithms (described in Gonz\'alez 1993) as
explained in Pedraz et al. (2002). In each case, the template that was
used was obtained as the mixture of 6 different spectral types (G5\,V,
G7\,V, K0\,III, K2\,III, M3\,III and M5\,III) which minimizes
intrinsic differences with the galaxy spectra. In order to avoid
systematic differences between indices at diferent spectral
resolutions, all the spectra were broadened up to the largest
$\sigma_0$ of the galaxy sample ($\sim 370$\,km\,s$^{-1}$). The
measured indices correspond to a central aperture of radius $R_{\rm
eff}$/8 (or 1\,arcsec for galaxies with $R_{\rm eff} < 8$\,arcsec) and
were corrected to the system defined by the models. See {\tt
http://www.ucm.es/info/Astrof/ellipt/CATRIPLET.html} for a database
with the indices and velocity dispersions.

Figure~\ref{I-sigma} shows the obtained relations of the indices and
the central velocity dispersions of the galaxies. Although, at first
sight, the two Ca\,{\sc ii} triplet indices do not follow a neat
linear behaviour with $\log \sigma_0$, the significance levels of
Spearman rank-order tests (see the labels) show that negative trends
with velocity dipersion are significant. This result is highly
surprising since it is the first evidence for an anti-correlation
between a metal-line index and the velocity dispersion. Note that
classical metallicity indicators in the blue spectral range increase
with $\sigma_0$ (e.g. Mg$_2$, $<$Fe$>$). In any case, it is worth
noting that the spread of CaT$^*$ and CaT values is only $\sim 5$ per
cent of the mean values. Probably, this is the reason why previous
works did not find significant variations of the Ca\,{\sc ii} triplet
in their galaxy samples. For the sTiO and MgI indices we find clear
increasing trends with $\log \sigma_0$, whereas we do not detect any
significant trend for the PaT index. Only NGC\,4742 significantly
departs from the mean PaT value, revealing a central young stellar
population in agreement with other determinations in the optical
(Gorgas et al. 1990; Trager 1997).


Although a linear fit is representative of the general behaviour as a
whole, different trends with velocity dispersion are apparent. While
the indices CaT$^*$, CaT and sTiO of low mass Es ($\log \sigma_0 \la
2.20$) are roughly independent of $\sigma_0$, galaxies with $2.30 \la
\log \sigma_0 \la 2.50$ depart from the above trend showing lower
(CaT$^*$ and CaT) and larger (sTiO) values. Also, some of the most
massive Es ($\log \sigma_0 \ga 2.50$) significantly deviate from the
fit attaining the typical indices values of low mass Es.


\section{Interpretation and discussion}
\label{discussion}

\begin{figure*}
\centerline{\hbox{
\psfig{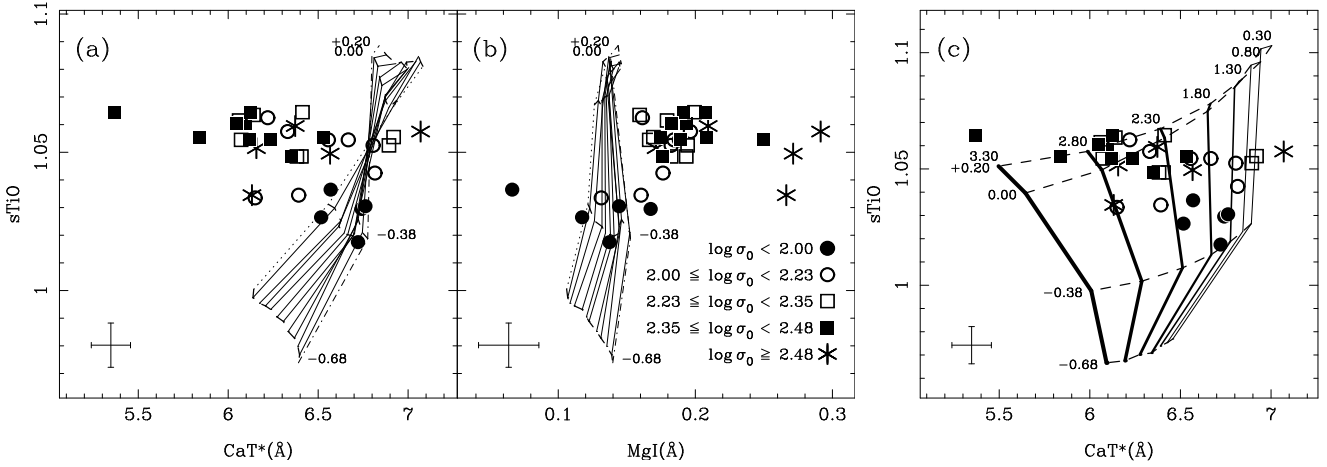}
}} 
\caption{\small 
CaT$^*$--sTiO and MgI--sTiO diagrams for the galaxy sample corrected
to the system defined by the models at 370\,km\,s$^{-1}$ spectral
resolution. Panels (a) and (b) show SSPs model predictions with fixed
Salpeter IMF slope ($\mu = 1.3$). Age varies from 5.01\,Gyr (dotted
line) to 17.78\,Gyr (dash-dotted line) with solid lines for
intermediate values ($\Delta\log$\,[age(Gyr)] $\sim
0.05$). Metallicity spans from $-0.68$ to $+0.20$ as in the labels
(dashed lines). Panel (c) shows SSPs model predictions at fixed age of
17.78\,Gyr with varying power-like IMF slope ($\mu = 0.3 - 3.3$, see
the labels) and metallicity (as in panel a). Different symbols
indicate galaxies within distinct ranges of central velocity
dispersion as it is shown in the key (b). Typical error bars for the
whole sample are given.}
\label{3panels}
\end{figure*}

To analyse the previous relations we make use of our SSP model
predictions (VAZ02; CEN02), 
which were
transformed to the spectral resolution of the data
(370\,km\,s$^{-1}$) using specific polynomials corresponding to their
own age, metallicity and IMF (see VAZ02).

Given that the time evolution of the near-IR indices is virtually null
for SSPs of all metallicities and ages $\ga 3$\,Gyr (VAZ02, CEN02), we
can consider that the age has a negligible effect in this spectral
range. Figure~\ref{3panels} shows the distribution of the galaxy
sample in the CaT$^*$--sTiO and MgI--sTiO planes, with symbols
indicating different ranges of $\sigma_0$. The insensitivity of the
indices to age is apparent in Figs.~\ref{3panels}a,b. From
Fig.~\ref{3panels}a, one immediately can notice that, while low-mass
Es (filled circles) can be roughly fitted by with SSPs of metallicity
below solar ($\sim -0.4$\,dex), no age-metallicity combination can
account for the low CaT$^*$ values of massive Es (filled squares). The
spread and location of the most massive Es (asterisks) will be further
discussed. Moreover, typical random errors in CaT$^*$ are not able to
explain the spread among Es, suggesting that other parameters must be
taken into account to explain the observed trend. In this sense, we
consider the existence of i) non-solar abundance ratios (hereafter
NSAR), and/or ii) systematic variations of the IMF, as possible
solutions to the above inconsistency.

NSAR could play an important role to interpret Fig.~\ref{3panels}a. In
fact, the high MgI values in massive Es (Fig.~\ref{3panels}b) confirm
the behaviour found for Mg in the optical. In this scenario, the
position of massive Es in Fig.~\ref{3panels}a implies that Ca should
be underabundant in these galaxies, as was already suggested by e.g.
Vazdekis et al. (1997) using the Ca4227 Lick/IDS index. As discussed
in VAZ02, the use of $\alpha$-enhanced scaled-solar isochrones
(Salasnich et al. 2000) predicts even larger CaT$^*$ values (by $\sim
0.5$\AA) for metal rich, old SSPs. Therefore, given that
overabundances of other $\alpha$-elements exist, and the contribution
of Ca to the total metallicity is lower than 0.5 percent, massive Es
should have extremely low abundances of Ca to account for the observed
values. Isochrones with accurate degrees of enhancement for the
different $\alpha$-elements as well as stellar spectral libraries with
appropriate element ratios are indeed necessary to tackle with the above
problem.

Could it be that massive Es are much more metal rich than the models,
i.e., much more than two times solar metallicity? In this case, the
shift of the red giant branch to lower temperatures would predict a
significant decrease in the CaT$^*$ and CaT values. However, one would
expect then sTiO values much larger than those observed (and also
higher Fe values in the optical), but the measured indices are in
reasonable agreement with the current model predictions.

\begin{figure}
\centerline{\hbox{
\psfig{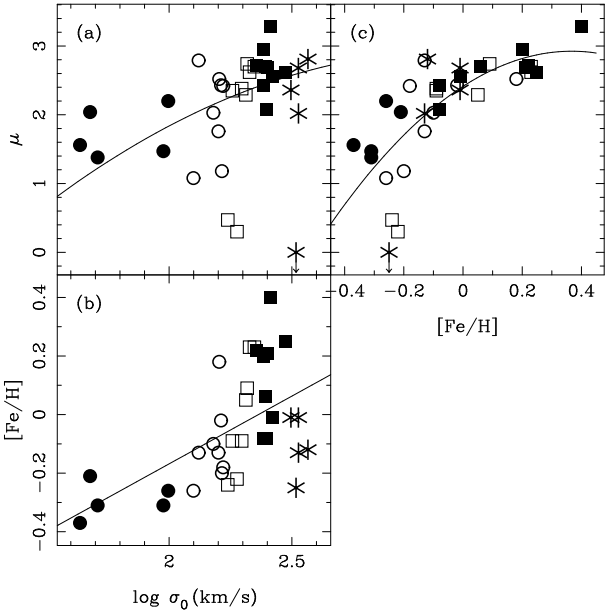}
}}
\caption{\small 
The $\mu$--[Fe/H]--$\log \sigma_0$ relation derived from
Fig.~\ref{3panels}c. Panels show the projection of the galaxies onto
the planes $\mu$--$\log \sigma_0$ (a), [Fe/H]--$\log \sigma_0$ (b),
and $\mu$--[Fe/H] (c). Lines indicate least-squares polynomial fits to
all data: (b) ${\rm [Fe/H]} = -1.09 + 0.46\;\log \sigma_0$; (c) $\mu =
2.41 + 2.78\;{\rm [Fe/H]} - 3.79\;{\rm [Fe/H]}^{2}$; (a) is straightly
derived from (b) and (c). Symbols are as in Fig.~\ref{3panels}.}
\label{FundTube}
\end{figure}

The alternative to interpret the data on the basis of the current
model predictions is to assume that the IMF may be varying among
Es. Since age can be considered a secondary parameter, in
Fig.~\ref{3panels}c we plot the SSPs model predictions for 17.78\,Gyr
(a different age does not qualitatively alter the final conclusions)
and different IMF slopes and metallicities. In this new scenario, the
data can be interpreted by means of IMF slope and metallicity
variations, suggesting that low-mass Es (filled circles) exhibit lower
metallicities and lower IMF slopes (lower dwarf/giants ratio) than
massive Es (filled squares). In the light of this interpretation,
reading $\mu$ and [Fe/H] values from the grid in Fig.~\ref{3panels}c
allows us to fit a one-parameter relation between $\mu$, [Fe/H] and
$\log \sigma_0$ for Es in the sense that, the larger $\sigma_0$, the
larger the metallicity and the IMF slope (Figure~\ref{FundTube}).
Note, however, that very massive Es (asterisks) deviate from the above
relation exhibiting lower metallicities and slightly flatter IMFs.
To guide the eye, solid lines represent simple polynomial fits to the
data on the $\mu$--[Fe/H] and [Fe/H]--$\log \sigma_0$ planes, whilst
the relation in the $\mu$--$\log \sigma_0$ plane is readily derived
from the two previous ones. Using Monte-Carlo simulations (following a
procedure similar to that of Kuntschner et al. 2001), we have checked
that this relation is not driven by a combined effect of Poisson noise
and non-orthogonal diagnostic diagrams, thus concluding that Es can
indeed be explained using $\mu$--[Fe/H]--$\log \sigma_0$ curves.

Although the universality of the IMF is a highly controversial
question (see e.g. Eisenhauer 2001 for a review), there are
theoretical arguments suggesting that the IMF of metal-rich
star-forming regions must be biased towards low mass stars due to a
more efficient cooling rate (e.g. Larson 1998). On this base, the
universal IMF by Padoan et al. (1997) --which depends on the physical
conditions of the star formation site-- predicts such a behaviour. The
observational $\mu$--[Fe/H] trend of Fig.~\ref{FundTube}c
qualitatively agrees with these theoretical arguments, but it still
needs to be related with the galactic mass (or $\log \sigma_0$).
A time-extended (non-instantaneous) star formation history or several
star-forming episodes could drive such a relation. If mergers and
accretion of other galaxies with pre-enriched gas have been frequent
phenomena during the evolution of massive Es, subsequent star
formation episodes should increase the metallicity and, as a
consequence, the IMF would gradually tend to produce a larger relative
number of low mass stars. In this case, the infall of non-primordial,
previously enriched, gas is needed to ensure the increasing
metallicity. Although an extended star formation could be enough to
explain the increasing metallicity, the addition of dissipative
hierarchical scenarios would emphasize such a behaviour.

This interpretation implicitly characterizes the IMF as a
time-dependent function (it gets steeper with time) in agreement with
several previous work (e.g. Larson 1998), suggesting that the IMF in
the early Universe could be biased towards very massive stars. A
prompt initial enrichment --straight consequence of a time-dependent
IMF-- could account for the observed $\alpha$-enhancements (Vazdekis
et al. 1996). This could also explain several keys like the G-dwarf
problem, the scarcity of Population III stars, and the high abundance
of heavy elements in the intergalactic medium of clusters of galaxies
(Larson 1998). In addition, the existence of a $\mu$--[Fe/H]--$\log
\sigma_0$ relation would also have important consequences for the
interpretation of the blue spectra of Es. In particular, the
$<$Fe$>$--$\log \sigma_0$, H$\beta$--$\log \sigma_0$, and the
mass--luminosity (in $B$ band) relations could only be reconciled by
introducing an age sequence in the sense that the larger $\log
\sigma_0$ the lower the mean age in the central parts (CEN02). Another
problem could be visual-infrared colors. For example, for a model with
$\mu = 2.8$, [Fe/H] $= +0.2$ and 12.6\,Gyr one expects $V-K = 3.52$
(Blakeslee et al. 2001).  This is just at the edge of the range of
observational values for Es 
(e.g., $V-K \la 3.50 \pm 0.05$ for Es in Frogel et al. 1978). 
However,
and again, a younger mean age for the central regions would help to
match the observed values ($V-K = 3.27$ for a model with $\mu = 2.8$,
[Fe/H] $= +0.2$ and 5.01\,Gyr).

The interpretation with a varying IMF actually revisits the classic
debate about the existence of a dwarf-enriched population in the
nuclei of Es, based on the strengths of the near-IR Na\,{\sc i}
doublet, the Ca\,{\sc ii} triplet and the FeH Wing-Ford band (Cohen
1978; Faber \& French 1980; Carter et al. 1986; Alloin \& Bica 1989;
Couture \& Hardy 1993). In particular, the weakness of the FeH band
was used as an argument against this possibility. Since these results
are mainly based on rather limited empirical synthesis models, they
are not strong enough to directly exclude the possibility of a
dwarf-heavy IMF. A proper calibration of its sensitivity to the
stellar parameters (in particular to metallicity; see Carter et
al. 1986) and its inclusion in modern stellar population models must
be performed before extracting any conclusion. 
In any case, since several model uncertainties affect the absolute
scale of the predicted index strengths (by $\sim$ 0.5\,\AA\ in the
case of CaT$^*$; see VAZ02), the derived $\mu$--[Fe/H]--$\log
\sigma_0$ relation must be considered on a relative basis.

Finally, we speculate on the very massive Es labelled with asterisks
in Figs.~\ref{3panels} and \ref{FundTube}. Whereas all (except one)
are boxy Es, with slow rotation and resolved cores, the rest of the
sample are mostly disky Es, fast rotators with power-law
cores. According to Faber et al. (1997), disky Es are consistent with
their formation in gas-rich mergers, whereas boxy Es could be the
by-products of gas-free stellar mergers. In the context of our
interpretation, this suggests that boxy Es should exhibit lower
metallicities (and, therefore, lower sTiO values) at the time that
keep a more primordial, flatter slope IMF than disky Es (thus leading
to larger values of CaT$^*$ and CaT), in agreement with what we
observe in Figs.~\ref{I-sigma} and ~\ref{FundTube}. Also, the fact
that boxy Es seem to be older than disky Es (de Jong \& Davies 1997;
Ryden, Forbes \& Terlevich 2001) favours the last interpretation.

We want to note that, at the time of submission of this letter, Saglia
et al. (2002) presented similar anti-correlations between Ca\,{\sc ii}
indices and the velocity dispersion, using another large sample of
high-quality spectra. Another paper by Falc\'on-Barroso et al. (2003)
has been submitted with such anti-correlations for bulges of spiral
galaxies.

To conclude, more work in the areas of SN yields, stellar interior
models and stellar libraries is needed to clarify whether it is
possible to explain the current discrepancy between Ca\,{\sc ii}
measurements and models using non-solar abundance ratios. Until this
is accomplished, and in the light of the present models, the only way
out is to advocate for variations in the IMF. By no means we think
that we are presenting strong evidence against its universality.
Furthermore, the scenario of a time-dependent IMF poses difficulties
to explain other observables. More modern, careful work using
indicators in the near-IR and the optical is needed to resolve the
issue of a non-standard IMF.

\section*{ACKNOWLEDGMENTS}

We are indebted to the anonymous referee for very useful
suggestions. This work was supported by the Spanish research project
No. AYA2000-977

\label{lastpage}

\end{document}